Differential activation patterns in the same brain region led to opposite emotional states

Kazuhisa Shibata[1,2], Takeo Watanabe[1,2], Mitsuo Kawato[1], and Yuka Sasaki[1,2]

[1] Brain Information Communication Research Laboratory Group, Advanced Telecommunications Research Institute International, 2-2-2 Hikaridai, Keihanna Science City, Kyoto 619-0288, Japan

[2] Department of Cognitive, Linguistics, & Psychological Sciences, Brown University, 190 Thayer Street, Providence, RI 02912, USA

Correspondence: Mitsuo Kawato (kawato@atr.jp)

**ABSTRACT**

In human studies, how averaged activation in a brain region relates to human behavior has been extensively investigated. This approach has led to the finding that positive and negative facial preferences are represented by different brain regions. However, using a multi-voxel pattern induction method we found that different patterns of neural activations within the cingulate cortex (CC) play roles in representing opposite emotional states. In the present study, while neutrally-preferred faces were presented, activation patterns in the CC that corresponded to higher (or lower) preference were repeatedly induced by the pattern induction method. As a result, previously neutrally-preferred faces became more (or less) preferred. We conclude that a different activation pattern in the CC, rather than averaged activation in a different area, represents and causally determines positive or negative facial preference. This new approach may reveal importance in an activation pattern within a brain region in many cognitive functions.

# INTRODUCTION

A traditional approach in human studies is to examine how averaged activation in a brain region relates to behavior. Results obtained by this approach led most theories of cognitive functions in the human brain to assume that a different region or a group of regions in the human brain plays a role in a different function. While this approach has greatly advanced the understandings of neural mechanisms of human cognitive functions, it cannot effectively reveal a differential role of a pattern of activity within the same region in a different cognitive function. In animal studies, the importance of a role of activity of a certain group of neurons in a region rather than mean activity of the region has been observed [1, 2]. Thus, to better understand neural mechanisms of human cognitive functions, it is necessary to investigate how a different pattern of activation within a region plays a different role.

Facial preferences influence a wide range of social outcomes from face perception to social behavior [3-13] and therefore has been a subject of great interests. Theories of facial preferences have also been developed with the general consensus that positive and negative facial preferences are represented by different brain regions [14-20].

A recently developed online functional magnetic resonance imaging (fMRI) multi-voxel pattern induction method has allowed us to induce a different pattern of activation within the same brain region [21]. In the present study, using this multi-voxel pattern induction method we tested whether a different pattern of activations within a single brain region can sufficiently change facial preferences in the positive or negative direction.

In the experiment, we chose the cingulate cortex (CC) as the target brain region for multi-voxel pattern induction since the CC was found to be the best region whose activation patterns represent both positive and negative facial preferences in the current study, among the regions previously implicated in facial preference [14-20]. We tested whether subjects' preferences to neutrally-preferred faces could be changed toward a positive (or negative) direction by the multi-voxel pattern induction method, which induced activation patterns in the CC that correspond to higher (or lower) preference with presentations of the neutrally-rated faces to generate a new association between the faces and manipulated preferences. As a result, the previously neutrally-rated faces became significantly more (or less) preferred. While subjects' facial preferences were successfully changed, subjects remained unaware of the aim to change their facial preferences. On the contrary to the previous belief that a different brain region plays a role in positive or negative facial preference, our results are in accord with the hypothesis that it is a different activation pattern in the CC that represents and causally determines positive or negative facial preference.

**RESULTS**

**Determination of the CC as a target region**

To determine a single region that would be used as a target for multi-voxel pattern induction, we first conducted a pilot experiment (see Fig 1A and S1 Fig, and *Pilot experiment* in Materials and Methods). Results of the pilot experiment showed that the CC most accurately reflects subjects' behavioral preference ratings both in the negative and positive directions among the regions previously implicated in facial preference [14-20] (S1A Fig). Thus, we determined the CC as the target region for multi-voxel pattern induction in the main experiment.

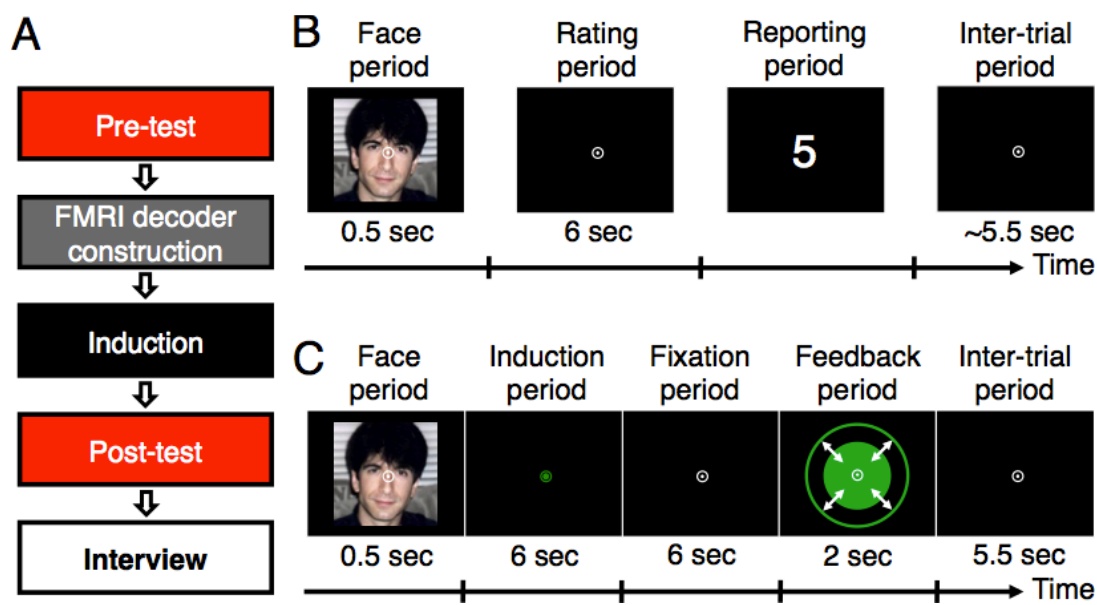

**Fig 1. Procedures of the experiments.** (A) Five stages of the main experiment. The pilot experiment consisted only of the first two stages. (B) Procedure of a trial in the pre-test, post-test, and fMRI decoder construction stages. (C) Procedure of a trial in the induction stage. The outer green circle in the feedback period indicates the maximum disk size.

**Main experiment with multi-voxel pattern induction**

This experimental design was made, with the aim to test whether induction of activation patterns in the CC which represent higher (or lower) preference with presentations of neutrally-preferred faces can cause these faces to be more (or less) preferred. The main experiment consisted of 5 stages (see Fig 1A and *Main experiment* in Materials and Methods for details): pre-test (1 day), fMRI decoder construction (1 day), induction (multi-voxel pattern induction, 3 days), post-test (20 minutes after the offset of the induction stage), and interview of subjects (immediately after the post-test stage). In the pre-test stage we measured a distribution of behavioral preference ratings to faces for each subject. In the fMRI decoder construction stage, for each subject we constructed a preference decoder to estimate the preference rating which was represented by the activation pattern in the target region (the CC). In the induction stage, multi-voxel pattern induction was administered by using the preference decoder. Through multi-voxel pattern induction, activation patterns in the CC were made to be similar to specific patterns, which represent higher (or lower) preference ratings, in association with presentations of neutrally-rated faces. In the post-test stage, subjects' behavioral preference ratings to the same faces as in the pre-test stage were measured so that we could test whether behavioral preference ratings to the previously neutrally-rated faces were changed due to induction of the specific activation patterns in the CC.

In the pre-test stage, subjects' behavioral preference ratings to 400 face pictures were measured. In each trial (Fig 1B), after the brief presentation of each face, subjects determined their preference to the face on a scale of 1-10 (1 for the lowest, 10 for the highest) during a 6-sec rating period. In a subsequent reporting period, they were asked to report the determined preference rating.

Based on the behavioral preference ratings recorded in the pre-test stage, we selected neutrally-rated faces which would be used in the subsequent stages. For each subject, one set of 15 neutrally-rated faces was randomly selected for use in the induction stage and called "induction faces". Another set of 15 neutrally-rated faces was also randomly selected as "baseline faces" for a control set which was not shown during the induction stage. Since both induction and baseline faces were neutrally-rated in the pre-test stage, the average behavioral ratings were the same between the two sets of faces in the pre-test stage. Thus, comparison of changes in the behavioral preference ratings between the induction and baseline faces in the post-test stage would indicate whether pairings of specific activation patterns in the CC with the induction faces during the induction stage are sufficient to change the behavioral preference ratings to the induction faces.

In the fMRI decoder construction stage, we constructed the preference decoder (sparse linear regression [22]), which would be used during the subsequent induction stage for the target region (the CC). In the fMRI decoder construction stage, subjects again conducted the preference-rating task in the fMRI scanner (see Fig 1B and *Main experiment* in Materials and Methods). Based on the fMRI signals measured during the rating period and corresponding behavioral preference ratings, for each subject we constructed the preference decoder to estimate the subject's behavioral preference ratings from activation patterns in the CC.

The purpose of the 3-day induction stage was to associate the induction faces with specific activation patterns in the CC that represent higher (or lower) preference ratings through the multi-voxel pattern induction method using the preference decoder. Subjects were randomly assigned to either a higher-preference ($N$=12) or lower-preference ($N$=12) group, but were not informed of their assigned group. Each trial consisted of face, induction, fixation, feedback, and inter-trial periods (Fig 1C). During the face period, subjects were presented with one of the induction faces. In the induction period, subjects were instructed to somehow regulate their brain activity to make the size of a solid green disk (presented in the subsequent feedback period) as large as possible. Subjects were encouraged to enlarge the disk size so that they would receive a payment bonus proportional to the mean disk size. Subjects were given no further instructions. The size of the disk presented in the feedback period served as a feedback signal and reflected an estimated preference rating from the CC, which was calculated by applying the preference decoder to the activation pattern of the CC obtained in the preceding induction period of the trial (see *Main experiment* in Materials and Methods for details). However, the computation of the disk size was opposite in its direction between the two groups, although the instructions given to the two groups were exactly the same. For the higher-preference group, the disk size was proportional to the estimated rating from the CC activation pattern. That is, if the CC activation became more similar to the patterns corresponding to higher preference, the disk size became larger. In contrast, for the lower-preference group, a lower estimated rating made the disk larger. This way made the instruction and the range of feedback signals to both groups identical. Note that all other information, including the intended preference direction, the purpose of the induction stage, and the meaning of the disk size was withheld from subjects so that knowledge of the purpose of the experiment would not influence subjects' rating criteria in the post-test stage.

**Changes in behavioral preference rating as a result of multi-voxel pattern induction**

To confirm that subjects' behavioral preference ratings to the induction faces changed as a result of associations of the induction faces with CC activation patterns of higher (higher-preference group) or lower (lower-preference group) preference ratings during the induction stage, the following three criteria have to be satisfied. First, subjects' preference ratings as behavioral measures for originally neutrally-rated induction faces must be significantly higher with the higher-preference group and lower with the lower-preference group in the post-test stage than in the pre-test stage. Second, to rule out the effect of mere exposure to faces on preferences to the faces [23], the subjects' behavioral preference ratings must be unchanged simply by repeated exposures to the faces during the fMRI decoder construction and induction stages. Thus, a new group of subjects as a control group ($N$=6) underwent an experiment in which visual presentations were identical to those for the higher- and lower-preference groups while no multi-voxel pattern induction was administered (see *Main experiment* in Materials and Methods for details). Third, changes in subjects' behavioral preference ratings after the induction stage must occur specifically for the induction faces, but not for the baseline faces, which were originally neutrally rated but were not used during the induction stage.

To test if the results of the main experiment met these criteria, a three-way mixed-model ANOVA with factors being test stage (pre- vs. post-test stage), face type (induction vs. baseline face), and group (higher-preference, lower-preference vs. control group) was applied to the behavioral preference ratings (Fig 2). The main effects of test stage ($F_{1,27}$ = 4.61, $P$ = 0.04) and group ($F_{2,27}$ = 3.56, $P$ = 0.04) were significant. Significant interactions were obtained between test stage and group ($F_{2,27}$ = 8.31, $P$ <

$10^{-2}$), between face type and group ($F_{2,27} = 10.84$, $P < 10^{-3}$), between test stage and face type ($F_{1,27} = 4.63$, $P = 0.04$), and among the three factors ($F_{2,27} = 13.10$, $P = 10^{-4}$). Post-hoc t-tests revealed that in the post-test stage subjects' behavioral preference ratings to the induction faces were significantly higher for the higher-preference group (Fig 2, red; paired two-tailed t-test, $t_{11} = 4.78$, $P < 10^{-3}$; Bonferroni corrected) and significantly lower for the lower-preference group (Fig 2, blue; $t_{11} = 3.31$, $P < 10^{-2}$, Bonferroni corrected) than in the pre-test stage. The results meet the first criterion. Moreover, post-hoc t-tests showed no significant changes in subjects' behavioral preference ratings between the two test stages for the control group (Fig 2, gray; $t_5 = 0.69$, $P = 0.52$), meeting the second criterion. For all of the 3 groups, no significant change in subjects' behavioral preference ratings was observed for the baseline faces which were neutrally rated in the pre-test stage but not presented during the induction stage, meeting the third criterion (Fig 2, baseline faces; $t_{11} = 1.15$, $P = 0.27$ for the higher-preference group; $t_{11} = 0.45$, $P = 0.66$ for the lower-preference group; $t_5 = 0.72$, $P = 0.50$ for the control group). From all of these results, we conclude that association of originally neutrally-rated faces with covert induction of activity patterns in the single brain region, the CC, led to changes in facial preference specifically for those faces, and in a specific preference (positive or negative) direction.

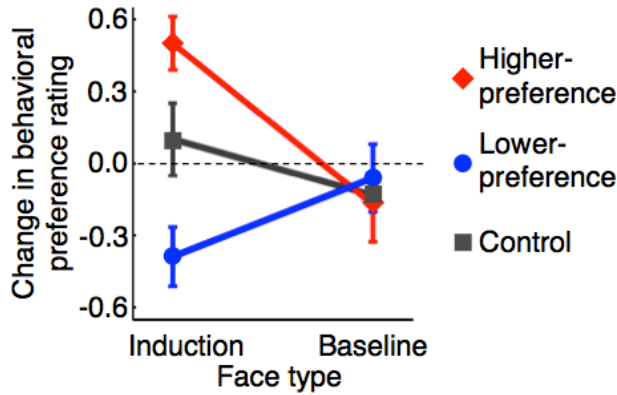

**Fig 2. Changes in the behavioral preference ratings.** The mean (± s.e.m) changes in subjects' behavioral preference ratings to the induction faces and baseline faces in the post-test stage in comparison to the pre-test stage. Red diamonds, blue circles, and gray squares represent the higher-preference ($N$=12), lower-preference ($N$=12), and control ($N$=6) groups, respectively.

**Subjects were unaware of what the disk size represented**

It is important that subjects were unaware of our manipulation and what the disk size represented, since knowledge or suspicion of what the disk size represented could have significantly influenced subjects' rating criteria. In the interview stage, which was held right after the post-test stage, subjects from the higher- and lower-preference groups were asked whether they knew or suspected what the disk size represented and what, if anything, they tried to do to increase the disk size. None of their responses indicated even a slightest understanding of the true workings of the experiment (see S1 Data for details). Subjects were then debriefed on how the disk size was computed, and asked to guess whether they had been assigned to the higher- or lower-preference group. The accuracies of their guesses were indistinguishable from chance for the higher-preference (Chi-square test, $\chi$ = 0.17, $P$ = 0.68) and lower-preference ($\chi$ = 0.00, $P$ = 1.00) groups (Fig 3). These results of the interview stage suggest that subjects remained unaware of what the disk size represented. That is, it was beyond subjects' will that induction of specific activation patterns in the single region changed facial preference in a specific (positive or negative) direction.

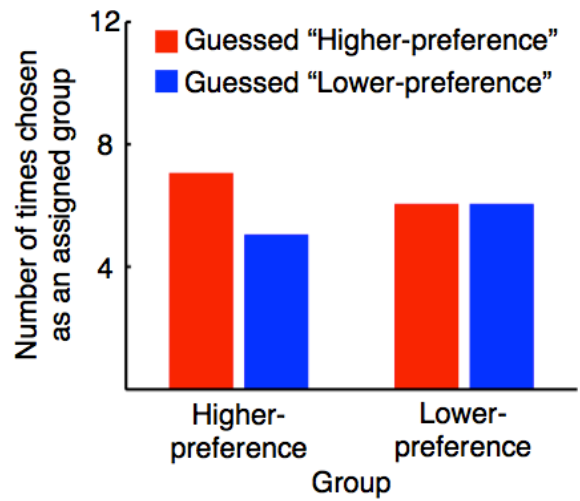

**Fig 3. Results of the interview stage.** The two left bars and two right bars show subjects' guesses as to which group they had been assigned for the higher- ($N$=12) and lower- ($N$=12) preference groups, respectively.

**Effects of inductions of a pattern of activation on the changes in the behavioral preference ratings**

The present study tested whether subjects' behavioral preference ratings to the neutrally-rated faces can be altered by repetitive inductions of a certain pattern of activation within the same area. We found that the behavioral preference ratings to the neutrally-rated faces were indeed changed in the post-test stage from the pre-test stage (Fig 2). If activation patterns in the CC during the induction stage caused the changes in the behavioral preference ratings, there should be a quantitative relationship between the induced shifts in activation patterns in the CC during the induction stage and the observed changes in subjects' behavioral preference ratings in the post-test stage.

To examine this relationship, we first quantified the activation patterns in the CC during the induction stage as described in (*1*), and called it the metric as *induced CC-activation shifts*. Then, we tested the prediction that the degree of the induced CC-activation shift should be correlated with the degree of the behavioral preference rating change as described in *(2)*.

***(1) Calculating CC-activation shifts during the induction stage.*** The CC-activation shift was calculated using the preference decoder in the following ways. As mentioned above, the preference-related component of an activation pattern in the CC in the induction stage was represented as an estimated preference rating by the preference decoder (see *Main experiment* in Materials and Methods for details). The preference decoder was constructed based on the individual behavioral preference ratings obtained during the fMRI decoder construction stage for each subject. The average behavioral preference rating varied with a different subject during the fMRI decoder construction stage. Therefore, to appropriately evaluate and compare the induced shifts in activation patterns in the CC during the induction stage across subjects or groups of subjects, we subtracted the above-mentioned average behavioral preference rating during the fMRI decoder construction stage from the estimated preference rating from the CC during the induction stage for each subject. The resultant value of the subtraction is called an induced CC-activation shift. That is, an induced CC-activation shift represents how far an induced activation pattern in the CC during the induction stage was shifted away from the activation pattern for the average behavioral preference rating for each subject (see *Main experiment* in Materials and Methods for detailed description of the induced CC-activation shift). For instance, a "0" induced CC-activation shift indicates that the activation pattern of the CC was biased in neither the positive nor negative preference direction.

If the induced shifts in the activation patterns in the CC during the induction stage led to the observed changes in the behavioral preference ratings between the pre- and post-test stages, the induced CC-activation shifts should have a quantitative relationship to the behavioral preference rating changes.

***(2) Testing of the prediction.*** The following results support the above-mentioned prediction that there is a quantitative relationship between induced CC-activation shifts and changes in the behavioral preference ratings. We found a significant correlation between the changes in subjects' behavioral preference rating and the induced CC-activation shifts averaged over the 3-day induction stage (Pearson's correlation test, $r_{22}$ = 0.78, $P < 10^{-4}$; Fig 4). Moreover, as described below the close-to-zero intercept and positive slope of the regression line (Fig 4) suggest a causal relationship of the induced CC-activation shifts to the changes in the behavioral preference ratings. The zero intercept of the regression line indicates that if an induced CC-activation shift was zero (in other words, activation patterns corresponding to higher or lower preference ratings did not occur in the CC), then no behavioral change occurred. The positive slope of the regression line means that if the induced CC-activation shift is positive (or negative), the behavioral change is also positive (or negative). Third, the slope was close to one. This indicates that the magnitude of the behavioral preference change was almost equivalent to that of the induced CC-activation shift. Furthermore, even if the group effect (higher- or lower-preference group) was removed by reversing the sign of the data for the lower-preference group (S2 Fig), a significant correlation was preserved ($r_{22}$ = 0.57, $P < 10^{-2}$). These results suggest a cause and effect relationship [24]; a different pattern of activation in the CC changed facial preference in a positive or negative direction.

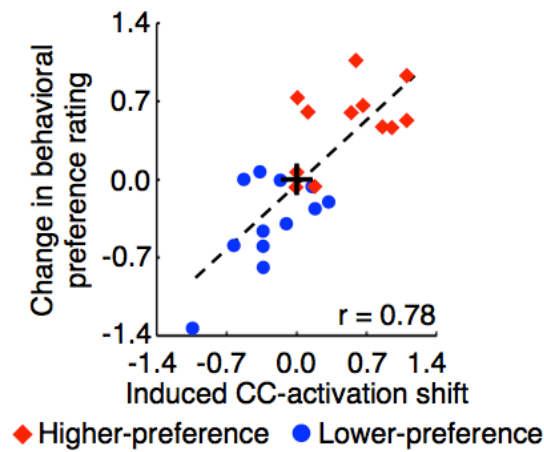

**Fig 4. Significant correlation between the induced CC-activation shifts and the degree of changes in the behavioral preference ratings.** Here shown is the scatter plot of the induced CC-activation shift during the 3-day induction stage vs. the change in subjects' behavioral preference rating for the higher- (red diamonds; $N$=12) and lower- (blue circles; $N$=12) preference groups. There was no significant outlier in the plot (Grubbs' test, $P > 0.05$). The black cross and broken line indicate the origin (0, 0) and the ordinary least-square regression, respectively.

**The differential activation patterns in the CC can explain the bi-directional changes in the behavioral preference rating while mere pairing of faces with monetary reward cannot account for the bi-directional behavioral changes**

We found significant bi-directional changes in the subjects' behavioral preference ratings to the induction faces in the post-test stage from the pre-test stage (Fig 2). In addition, the results of the quantitative analysis (Fig 4 and S2 Fig) suggested that a different activation pattern induced in the CC during the induction stage determined a direction and degree of the changes in subjects' behavioral preference ratings to the induction faces. However, is there any possibility that pairings of the induction faces with the monetary reward during the induction stage directly led to the subjects' behavioral preference rating changes in the post-test stage without relying on the differential activation patterns induced in the CC? This possibility arises since payment bonus (monetary reward) was given to subjects in proportion to the size of the green feedback disk during the induction stage. However, the following lines of evidence do not support this possibility.

First, the lack of difference in the amounts of monetary reward between the higher- and lower-preference groups is inconsistent with the model that the amount of reward should simply increase the preference. Assume that monetary reward was a direct deterministic factor to make faces more preferred. Then the amount of the payment bonus given to the higher-preference group should have been larger than the lower-preference group, because the behavioral preference rating increased in the higher-preference group and decreased in the lower-preference group (Fig 2). However, there was no significant difference in the amounts of the payment bonus between the higher-and lower-preference groups (Fig 5A; two-sample two-tailed t-test, $t_{22}$ = 0.19, $P$ = 0.85). Second, if monetary reward was a direct deterministic factor to make faces more preferred, the amount of the payment bonus should have been correlated with the degree of the behavioral preference rating change across the two groups. However, no significant correlation was found between them (Fig 5B; $r_{22}$ = 0.15, $P$ = 0.49). Third, if hypothesis for the direct role of reward is true, then the amount of the payment bonus should have been positively correlated with the estimated preference rating by the decoder from the CC activation patterns in each of higher- and lower-preference groups. Indeed, the estimated rating from the CC activation pattern was positively correlated with the amount of the payment bonus in the higher-preference group (Fig 5C; red diamonds). However, the estimated rating was *negatively* correlated with the amount of payment bonus in the lower-preference group (Fig 5C; blue circle). These correlations reflect the experimental procedure for the induction stage (see *Main experiment* in Materials and Methods). These lines of evidence clearly deny the possibility that monetary reward during the induction stage was the direct factor to induce both positive and negative preferences. They also support the conclusion that it is the differential activation patterns induced in the CC that led to the bi-directional changes in the behavioral preference ratings.

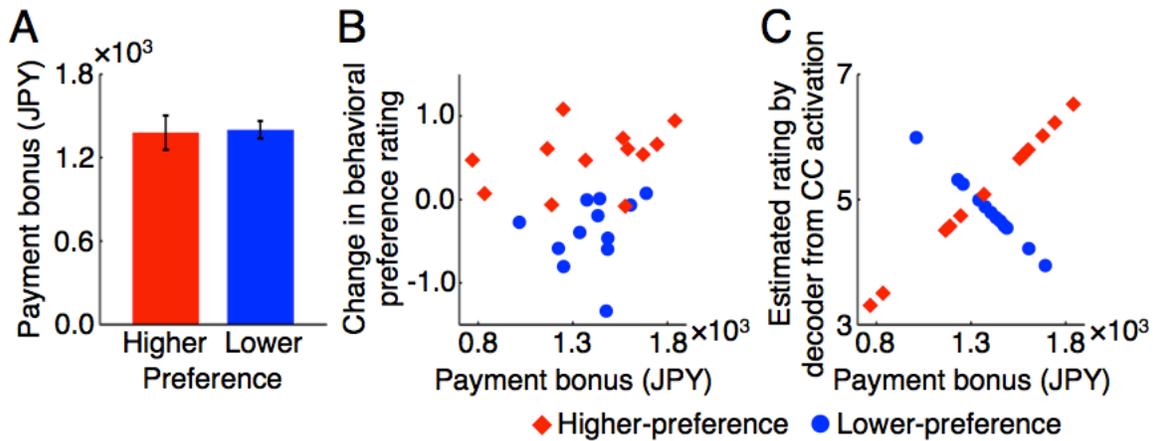

**Fig 5. Monetary reward, activation patterns of the CC during the induction stage, and the behavioral preference rating changes.** (A) The mean (± s.e.m) amounts of payment bonus (monetary reward) for the higher- (red; *N*=12) and lower- (blue; *N*=12) preference groups. (B) Scatter plot of the amount of payment bonus vs. the change in subjects' behavioral preference rating for the higher- (red diamonds) and lower- (blue circles) preference groups. (C) Scatter plot of the amount of payment bonus vs. the estimated rating from activation patterns of the CC during the 3-day induction stage for the higher- (red diamonds) and lower- (blue circles) preference groups.

**Induction of the preference-related activation patterns was largely confined to the CC**

The present results suggest that the different activation patterns in the CC cause different emotional states. How can we know that the CC mainly determines the emotional states? The size of the feedback disk provided to subjects in the induction stage was based on activation patterns only in the CC. However, this procedure alone does not assure that inductions of the preference-related activation patterns were confined to the CC. It is possible that, in concert with the successful induction of specific activation patterns in the CC, similar activation patterns occurred in some other regions during the induction stage, which might also contribute to the behavioral preference rating changes. To test the possibility that preference-related activation patterns occurred in any other region, we anatomically divided the brain into a total of 38 regions (the CC and 37 other regions) and analyzed activation patterns during the induction stage in each of the 38 regions (see *Leak test* in Materials and Methods for details) as described below.

If the activation patterns in the CC that represented higher (or lower) preference ratings during the induction stage "leaked out" and induced similar preference-related activation patterns in other regions, the activation patterns in the other brain regions than the CC should reconstruct the estimated ratings from the CC activation patterns during the induction stage on a trial-by-trial basis. To test this possibility, the estimated ratings from the CC activation patterns were reconstructed using activation patterns measured in each of the aforementioned 37 other brain regions during the induction stage, as well as the CC itself as a control. Reconstruction performance was defined as a correlation coefficient between the reconstructed values and the estimated ratings from the CC activation patterns (see *Leak test* in Materials and Methods for details). If there was leakage of the CC activation patterns to the other brain regions, the correlation coefficient should be high.

The Fisher-transformed correlation coefficients for the other 37 regions (Fig 6, gray) were significantly and markedly smaller than that for the CC itself (Fig 6, red; paired two-tailed t-test, $t_{23} > 14.50$, $P < 10^{-12}$, Bonferroni corrected). If the activation patterns in other brain regions had been closely linked to those in the CC, the correlation coefficients should have been as high as the CC. Thus, these results indicate that it is unlikely that the activation patterns induced in the CC leaked out to other regions. They also indicate that the CC mainly contributed to the behavioral preference rating changes in the current study.

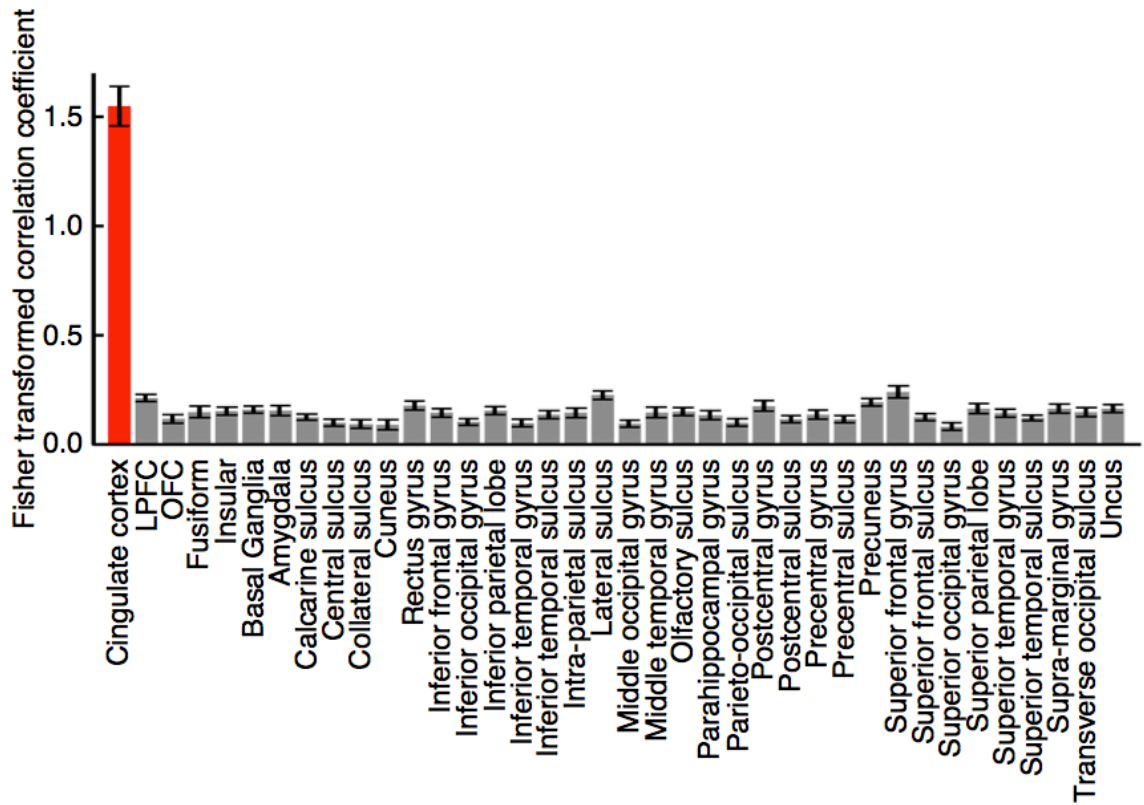

**Fig 6. Results of the leak test.** Each bar represents the mean (± s.e.m) Fisher-transformed correlation coefficients between the estimated ratings from the CC activation patterns during the induction stage and the reconstructed values of the estimated ratings from activation patterns in each of the CC (red) and other regions (gray). Subjects from the higher- and lower-preference groups were combined (*N*=24).

**DISCUSSION**

To our knowledge, this is the first study in which manipulation of a different brain activation pattern in a single region successfully changed facial preferences in the positive or negative direction. Previous neuroscientific approaches have revealed that a different region or a group of regions is involved in positive or negative facial preference [14-20] and developed the general consensus that positive and negative facial preferences are represented by different brain regions. Contrary to this general consensus, the present results using the multi-voxel pattern induction method [21] indicated a case in which highly selective activity patterns in the single brain region represent and causally determine both positive and negative facial preferences.

Additionally, our control analyses provide two important insights. First, the facial preference was not necessarily increased by a larger amount of monetary reward. Especially in the lower-preference group, the higher the monetary reward was, the less liked faces were. Second, when the CC was solely manipulated by the multi-voxel pattern induction method, the effect of the manipulation on other brain regions was minimal or ignorable. Together, the present results suggest that the induced activation pattern within the CC was involved in a causal role in changing the facial preference.

It has been found that a number of brain regions are involved in facial preferences [14-20]. Why was the CC in particular selected as the target region for multi-voxel pattern induction to change subjects' facial preferences? An important purpose of the current study was to test whether a different pattern of activity in a single brain region changes facial preference in an opposite direction. In the aforementioned pilot experiment, we found that the CC is the best region that codes *both* positive and negative facial preferences depending on the pattern of activation within the CC. Although the prefrontal cortex plays an important role in facial preferences in studies using a univariate analysis, it has been found that a different part of the prefrontal cortex is involved in each of positive and negative facial preference coding [18]. Thus, the criterion for the selection of the CC was not based on whether the CC is the most important region for facial preference, but based on the fact that the CC was found to code *both* positive and negative facial preferences in the current study.

In summary, the results of the present study indicate that highly selective activity patterns for higher or lower preference within the CC that were repeatedly paired with facial stimuli caused changes in subjects' facial preferences. Although these results do not deny important roles of other regions in facial preference, our study clearly demonstrates that inductions of different patterns of activation within the CC determine changes of facial preference in opposite directions beyond subjects' will [25].

**MATERIALS AND METHODS**

**Subjects**

Thirty-three naïve subjects (19 to 29 years old; 23 males and 10 females) with normal or corrected-to-normal vision participated in the study. All experiments and data analyses were conducted in the Advanced Telecommunications Institutes International (ATR). The study was approved by the Institutional Review Board of ATR in which the current study was conducted. All subjects gave written informed consents.

**Pilot experiment**

The purpose of the pilot experiment was to determine a target region of interest (ROI) to be used in the main experiment. Three subjects participated in the pilot experiment. The complete experiment consisted of two stages: pre-test (1 day) and fMRI decoder construction (1 day). The two stages were separated by at least 24 hours.

Only behavioral data were collected from the pre-test stage, during which subjects' behavioral preference ratings to face pictures were measured. Subjects performed a preference-rating task (Fig 1B) for a total of 400 trials across 20 runs. During each run, subjects were asked to fixate on a white bull's eye presented at the center of the display. A brief break period was provided after each run upon a subject's request.

We used a pool of 400 face pictures (200 males, 200 females, of a variety of races and ages) collected from several open databases [26-31]. A stimulus primarily consisted of a face, and usually included some body parts including hair, a neck and shoulders, as well as a background scene. Each face picture was 4° square in size. The order of presentation of faces was randomized for each subject.

Each trial (Fig 1B) consisted of a face period (0.5 sec), a rating period (6 sec), and a reporting period, in the respective order. During the face period, a face picture was presented for 0.5 sec at the center of the display. During the rating period, only a fixation point was presented at the center. Subjects were instructed to rate their preference to the previously presented face on a 10-point scale (1 for the lowest preference, 10 for the highest preference). During the reporting period, subjects were asked to report the preference rating by pressing two buttons (left and right) on a keyboard using the index and middle fingers of their right hand. At the beginning of the reporting period, a random number from 1 to 10 was selected and presented at the center of the display. Subjects were asked to adjust its value to their preference rating, pressing the left button to increment it. Values were "wrapped" so that when subjects attempted to increment past a value of 10, values would start over at 1. Subjects completed the reporting period by pressing the right button. After completion of the reporting period, the next trial began.

For each subject, 240 of the 400 face pictures were selected to be used for the subsequent fMRI decoder construction stage. Selections for pictures were based on subjects' individual behavioral preference ratings in the pre-test stage: the 100 highest-rated faces, the 100 lowest-rated faces, and 40 neutrally-rated faces.

In the fMRI decoder construction stage, we measured subjects' blood-oxygen-level-dependent (BOLD) signal patterns (see *MRI measurements and parameters*) while they once again conducted the preference-rating task on the 240 face pictures selected from the pre-test stage (the 100 highest-rated faces, the 100 lowest-rated faces, and 40 neutrally-rated faces). The measured BOLD signal patterns and behavioral preference ratings were in turn used to compute parameters for a preference decoder for each of different ROIs (see below). Task procedures were identical to those of the pre-test stage, except that an inter-trial period was added to the end of each trial, in which only a white fixation point was presented at the center of the display (Fig 1B). Subjects were asked to report their ratings within the reporting period (maximum of 5.5 sec). The duration of the inter-trial period varied across trials, depending on subjects' reporting time, so that the total duration of the reporting and inter-trial periods would be equal to 5.5 sec.

Each fMRI run for the fMRI decoder construction stage consisted of 20 task trials (1 trial = 12 sec), plus a 10-sec fixation period before the trials and a 2-sec fixation period after the trials (1 run = 252 sec). The fMRI data for the initial 10 sec were discarded to allow the longitudinal magnetization to reach equilibrium. Subjects conducted a total of 240 trials in 12 fMRI runs. Throughout each fMRI run, subjects were asked to fixate on a white bull's eye presented at the center of the display. A brief break period was provided after each fMRI run upon a subject's request.

Recorded fMRI data were preprocessed using the BrainVoyager QX software [32]. All functional images underwent 3D motion correction. No spatial or temporal smoothing was applied. Rigid-body transformations were performed to align the functional images to the structural image for each subject. A gray matter mask was used to extract BOLD signals only from gray matter voxels for further analysis.

We specified seven ROIs implicated in facial preference [14-20] according to anatomical data for each subject: the cingulate cortex (CC), lateral prefrontal cortex (LPFC), orbitofrontal cortex (OFC), fusiform area, insular cortex, basal ganglia, and amygdala. LPFC was defined as the middle frontal gyrus plus the inferior frontal sulcus. OFC was defined as the orbital gyrus plus the orbital sulci. The fusiform area was defined as the lateral occipitotemporal gyrus, the medial occipitotemporal gyrus, plus the occipitotemporal sulcus. The basal ganglia were defined as the caudate, the pallidum, the putamen, plus the nucleus accumbens. Voxels from the left and right hemispheres were merged for each ROI. The cortical regions were specified using an atlas on the BrainVoyager QX software [32]. A cortical surface for each subject was spatially normalized into a standard cortical surface using a cortex-based alignment method [33]. Then, the specified regions were projected into a native space for each subject. The subcortical regions were specified for each subject using an automated brain parcellation method [34] on the Freesurfer software (http://surfer.nmr.mgh.harvard.edu).

A time-course of BOLD signal intensities was extracted from each voxel in each ROI and shifted by 4 sec to account for the hemodynamic delay using the Matlab software. A linear trend was removed from the time-course. The time-course was z-score normalized for each voxel using all time points except for those for the initial 10 sec in each fMRI run to minimize baseline differences across the fMRI runs. The data samples for computing the decoder were created by averaging the BOLD signal intensities of each voxel for 3 volumes corresponding to the 6-sec rating period.

To construct a preference decoder for each ROI, we used a sparse linear regression algorithm [22], which automatically selected the relevant voxels within a ROI for decoding. Note that the behavioral preference ratings measured in this study were non-linear. Although they ranged from 1 to 10, preference measurement on the Likert-type scale cannot be considered strictly linear. Thus, before applying the sparse linear regression for each ROI, the behavioral preference ratings were linearized using an arc hyperbolic tangent function. An estimated rating $R_{decoded}$, that is the decoder output calculated based on an activation pattern for a trial, was obtained in each ROI by

$$R_{decoded} = W_{voxel}^T \cdot A_{voxel} + b.$$

Here, $A_{voxel}$ represents the activation pattern of voxels in the ROI for the trial. $W_{voxel}$ indicates linear weights for the voxels which were optimized by the sparse linear regression algorithm based on fMRI data which was used for training the decoder. $b$ corresponds to the decoder's constant term, which was determined for each subject as his/her average behavioral preference rating across all faces in the preference-rating task during the fMRI decoder construction stage. $A_{voxel}$ and $W_{voxel}$ are denoted as $n$-dimensional column vectors with $n$ as the number of voxels in each ROI. $T$ denotes matrix transpose. The inputs to the decoder were subjects' moment-to-moment brain activations in each ROI, while the outputs from the decoder represented the decoder's best estimate of the corresponding behavioral preference ratings.

Decoder performance for each ROI was defined as the correlation coefficient between actual subjects' behavioral preference ratings in the preference-rating task and the estimated ratings calculated from activation patterns of the ROI and evaluated by a leave-one-run-out cross validation procedure. In the cross-validation procedure, the pairs of the actual subjects' behavioral preference ratings and the activation patterns for the ROI measured on one fMRI run were treated as the test data (20 samples), while those measured on the remaining runs (220 samples) were used for training the decoder to estimate subjects' trial-by-trial behavioral preference ratings. Thus, 12 cross-validation sets were generated per subject. For each voxel, activation amplitudes of the training and test data were normalized by mean and variance of activation amplitudes of the training data so that mean and variance of voxel activation amplitudes would be zero and 1, respectively. The correlation coefficients for each ROI were first standardized using Fisher's transformation, averaged over the cross-validation sets, and then averaged across subjects, as shown in S1A Fig.

The result of the fMRI decoder construction stage showed that the highest decoder performance was obtained from the CC (S1A Fig). Consistent with this result, previous neuroimaging studies have reported that the CC is highly involved in facial preference [15, 18, 35] as well as preferential decision-making in general [36-38]. Thus, we selected the CC as the target region for the main experiment. Note that the highest decoder performance was also found in the CC when we evaluated decoder performance in the same way using the fMRI signals obtained in the fMRI decoder construction stage of the main experiment (S1B Fig). These results indicate robustness of the tendency in which the CC most accurately reflects subjects' facial preference in the preference-rating task in this study.

**Main experiment**

Thirty subjects participated in the main experiment. The main experiment consisted of five stages: pre-test (1 day), fMRI decoder construction (1 day), induction (multi-voxel pattern induction, 3 days), post-test (20 minutes after the induction stage), and interview (immediately after the post-test stage) stages, in this order (Fig 1A). The pre-test, fMRI decoder construction, and induction stages were separated by at least 24 hours. Thirty subjects in the main experiment were randomly assigned to one of the higher-preference ($N$=12), lower-preference ($N$=12), and control ($N$=6) groups. They were not informed about their assigned group.

The procedures of the pre- and post-test stages in the main experiment were identical to those of the pre-test stage in the pilot experiment. As in the pilot experiment, the 100 highest-rated faces, the 100 lowest-rated faces, and 40 neutrally-rated faces were selected to be used in the subsequent fMRI decoder construction stage. In addition, out of the 40 neutrally-rated faces, 15 were randomly selected for use in the induction stage ("induction faces") and another set of 15 was also randomly selected as a preference-matched control against the induction faces ("baseline faces"). The baseline faces were not shown during the subsequent induction stage.

For the higher- and lower-preference groups, the procedures of the fMRI decoder construction stage in the main experiment were identical to those in the pilot experiment (see *Pilot experiment*). A preference decoder for the CC was computed for each subject for use in the subsequent induction stage. To train the decoder, we used 240 data samples obtained from the 240 trials in the 12 fMRI runs. For each voxel, activation amplitudes of the training data were normalized by the mean and the variance of activation amplitudes of the training data so that the mean and the variance of voxel activation amplitudes would be zero and 1, respectively. The mean (± s.e.m) numbers of voxels selected by the sparse linear regression algorithm to decode the subjects' preference ratings were 219.8 ± 0.5.

For the control group, the visual presentations in the fMRI decoder construction stage were identical to those in the pilot experiment while the experiment was conducted outside the MRI scanner without fMRI measurements.

In the induction stage, which consisted of 3 daily sessions, subjects from the higher- and lower-preference groups were instructed to regulate the activation of their brains which were controlled by an online fMRI technique [21, 39-41]. On each day, subjects participated in up to 12 fMRI runs. The mean (± s.e.m) number of runs per day was 10.8 ± 0.2. Each fMRI run for the induction stage consisted of 15 trials (1 trial = 20 sec) preceded by a 30-sec fixation period (1 run = 330 sec). The fMRI data for the initial 10 sec were discarded to allow the longitudinal magnetization to reach equilibrium. During each run, subjects were instructed to fixate on a white bull's eye presented at the center of the display. After each fMRI run, a brief break period was provided upon a subject's request.

Each trial in the induction stage (Fig 1C) consisted of a face period (0.5 sec), an induction period (6 sec), a fixation period (6 sec), a feedback period (2 sec), and an inter-trial period (5.5 sec), in that order. During the face period, one of the 15 induction faces described above was presented for 0.5 sec at the center of the display. The order of presentation of the 15 induction faces was randomized for each fMRI run. During the induction period, the color of the fixation point changed from white to green, and no visual stimulus except for the fixation point was presented. Subjects were instructed to regulate activation of their brain, with the goal of making the size of a solid green disk presented in the later feedback period as large as possible. The experimenters provided no further instructions or strategies. During the fixation period, subjects were asked simply to fixate on the central white point. This period was inserted between the induction period and the feedback period to compensate for the known hemodynamic delay, which we assumed lasted 4 sec, during which activation patterns in the CC were calculated in time for a green disk to be shown in the subsequent feedback period. The feedback period presented the green disk for 2 sec. The size of the disk was determined based on the estimated rating (see below), which is the decoder output value, based on the BOLD signal pattern of the CC measured in the prior induction period. The green disk was always enclosed by a larger green concentric circle (10º in diameter), which indicated the disk's possible maximum size. The feedback period was followed by the inter-trial period, during which subjects were asked to fixate on the central white point. This period was followed by the start of the next trial.

The size of the disk presented during the feedback period was based on the estimated rating from the CC activation pattern, which was computed during the fixation period, as follows. First, measured functional images underwent 3D motion correction using the Turbo BrainVoyager software. Second, a time-course of BOLD signal intensities was extracted from each of the voxels in the CC identified in the fMRI decoder construction stage, and was shifted by 4 sec to account for the hemodynamic delay. Third, a linear trend was removed from the time-course for each voxel using a linear regression algorithm based on all time points except for those for the initial 10 sec in each fMRI run, and the BOLD signal time-course was z-score normalized for each voxel using BOLD signal intensities measured for 20 sec starting from 10 sec after the onset of each fMRI run. Fourth, the data sample used to calculate the size of the disk was created by averaging the BOLD signal intensities of each voxel for 3 volumes corresponding to the 6-sec induction period. Finally, the estimated rating was calculated from the data sample using the decoder constructed in the fMRI decoder construction stage. For the higher-preference group, the size of the disk was proportional to the estimated rating (ranging from 1 to 10). For the lower-preference group, the size of the disk was proportional to 11 minus the estimated rating so that a lower estimated rating resulted in a larger disk. In addition to the fixed amount of the compensation for participation in the experiment, a bonus of up to 3,000 JPY was paid to subjects based on the mean size of the disk during each day.

Induced CC-activation shifts shown in Fig 4 and S2 Fig indicate how far the activation patterns of the CC during the induction stage are from the activation patterns in the CC corresponding to the average behavioral preference rating. The induced CC-activation shift was calculated as follows. First, the estimated rating $R_{decoded}$ from an activation pattern of the CC for a trial was computed by the same method as in the pilot experiment, but only from the CC. That is, $R_{decoded}$ was computed by

$$R_{decoded} = W_{voxel}^T \cdot A_{voxel} + b.$$

Here, $A_{voxel}$ represents the activation pattern of voxels in the CC in the induction period. $W_{voxel}$ indicates linear weights for the voxels which had been computed for each subject in the fMRI decoder construction stage. $b$ corresponds to the decoder's constant term, and had been determined for each subject as her/his average behavioral preference rating in the preference-rating task during the fMRI decoder construction stage. This constant term varied across subjects. $A_{voxel}$ and $W_{voxel}$ are denoted as $n$-dimensional column vectors with $n$ as the number of voxels in the CC. $T$ denotes matrix transpose.

The induced CC-activation shift for a trial was defined by

$$R_{decoded} - b = W_{voxel}^{T} \cdot A_{voxel}.$$

Based on the following computation, the induced CC-activation shift represents how far the activation patterns in the CC during the induction stage is away from the activation pattern in the CC corresponding to the average behavioral preference rating initially for each subject. As described above, the constant term *b* was determined for each subject as her/his average behavioral preference rating in the preference-rating task during the fMRI decoder construction stage. The set of 240 faces (the 100 highest-rated faces, the 100 lowest-rated faces, and 40 neutrally-rated faces) used in the fMRI decoder construction stage were selected for each subject according to her/his behavioral preference ratings to the 400 faces presented in the pre-test stage. That is, the constant term *b* represents the subject's average behavioral preference rating for the population of faces. Thus, the induced CC-activation shift, which is calculated by subtracting the constant term *b* from the estimated rating $R_{decoded}$, represents how far the activation pattern in the CC is from the activation patterns in the CC corresponding to the average behavioral preference rating for each subject.

It is necessary to obtain the induced CC-activation shift in order to appropriately evaluate and compare induced activation patterns in the CC across subjects and groups during the induction stage. Note that an induced CC-activation shift being "0" indicates that the activation pattern of the CC was biased in neither the high nor low preference direction. A positive (or negative) value of the induced CC-activation shift indicates that the activation pattern of the CC was biased toward a positive (or negative) preference direction, compared to the activation pattern corresponding to the average behavioral preference rating for each subject. Thus, when the mean induced CC-activation shift is significantly higher (or lower) than 0, this means that subjects accomplished significant learning to induce the preference-related activation patterns in the CC that correspond to higher (or lower) preference rating.

With the control group, during the induction stage the induction faces were presented in the same way as with the higher- and lower-preference groups. On the other hand, unlike with the higher- and lower-preference groups, the experiment was conducted outside the MRI scanner without fMRI measurements for the control groups. Subjects from the control group conducted a fixation task (see below), instead of the task given to those from the higher- and lower-preference groups during the induction stage.

In the fixation task for the control group in the "induction" stage, during the 6-sec "induction" period, the luminance of the central fixation point slightly decreased (from green to dark green), returning to its original luminance 300 ms later. This luminance change occurred several times in an unpredictable manner during the 6-sec period. Subjects from the control group were asked to count the number of luminance changes and report whether the number of the changes was even or odd by pressing one of two buttons using the index or middle finger of their right hand during the fixation period. The task difficulty was controlled by using an adaptive staircase method, so that the overall task difficulty was kept constant throughout the induction stage; the degree of fixation luminance change was slightly increased in the trial following an incorrect answer, and slightly decreased after two consecutive correct answers. Otherwise, luminance was kept around the same. The mean (± s.e.m) task accuracy for the fixation task was 67.4 ± 5% across subjects. The green "feedback" disk was presented during the 2-sec "feedback" period. The size of the disk was determined randomly for each trial. Subjects were instructed to fixate on the center of the display during the feedback period.

**Leak test**

To test whether activations in any other region than the CC could reconstruct the estimated ratings from the CC activation patterns, we first anatomically divided the brain into a total of 38 regions (the CC and 37 other regions). The 37 other regions were specified using an atlas on the BrainVoyager QX software [32] and shown in Fig 6.

Second, for each subject we used the sparse linear regression algorithm [22] in attempt to reconstruct the estimated ratings from the CC activation patterns, from activation patterns measured in each of the aforementioned 37 other brain regions during the induction stage, as well as the CC itself as a control. A reconstructed value $R_{reconstructed}$ for a trial in each region was obtained by

$$R_{reconstructed} = W_{voxel}^T \cdot A_{voxel} + b.$$

Here, $A_{voxel}$ represents an activation pattern of voxels in a region for a trial. $W_{voxel}$ indicates linear weights for the voxels which were optimized for each subject by the sparse linear regression algorithm, and $b$ corresponds to the constant term, which was determined as the average estimated rating from the CC activation pattern during the induction stage for each subject. Reconstruction performance for the 38 regions was defined as a correlation coefficient between the reconstructed values and the estimated ratings from the CC activation patterns and evaluated by a leave-one-run-out cross-validation procedure (Fig 6).

**Apparatus**

Visual stimuli were presented on a LCD display (1024 × 768 resolution, 60 Hz refresh rate) during the pre- and post-test stages and via a LCD projector (1024 × 768 resolution, 60 Hz refresh rate) during fMRI measurements in a dim room. All visual stimuli were made using the Matlab software and Psychtoolbox 3 [42] on Mac OS X.

**MRI measurements and parameters**

Subjects were scanned in a 3T MR scanner (Siemens MAGNETOM Verio) with a 12-channel head matrix coil in the ATR Brain Activation Imaging Center. FMRI signals were measured using a gradient echo-planar imaging sequence. In the fMRI experiments, 33 contiguous slices (repetition time = 2 sec, voxel size = 3 × 3 × 3.5 mm$^3$, 0 mm slice gap, field of view = 192 mm × 192 mm, echo time = 26 ms, matrix size = 64 × 64, bandwidth = 2367 Hz/pixel, phase encoding direction: from anterior to posterior, slice order: interleaved) oriented parallel to the AC-PC plane were acquired, covering the entire brain. For an automated parcellation method [34], T1-weighted MR images (magnetization-prepared rapid gradient-echo or MP-RAGE; 256 slices, the number of

partition = 208, voxel size = 1 × 1 × 1 mm³, 0 mm slice gap, repetition time = 2250 ms, inversion time = 900 ms, echo time = 3.06 ms, flip angle = 9 deg, field of view = 256 mm, matrix size = 256 × 256, bandwidth = 230 Hz/pixel, phase encoding direction: from anterior to posterior, partition (2nd phase) encoding direction: from right to left) were also acquired during the fMRI decoder construction stage.

## ACKNOWLEDGEMENTS

We thank J. Dobres for editing an early draft, B. Seymour for helpful comments on the work, Y. Oshima and ATR BAIC for technical assistances.

## AUTHOR CONTRIBUTIONS

All the authors conceived and designed the study. K.S. collected data. K.S., Y.S. and M.K. analyzed data. All the authors discussed the results, wrote the manuscript.

# SUPPORTING INFORMATION

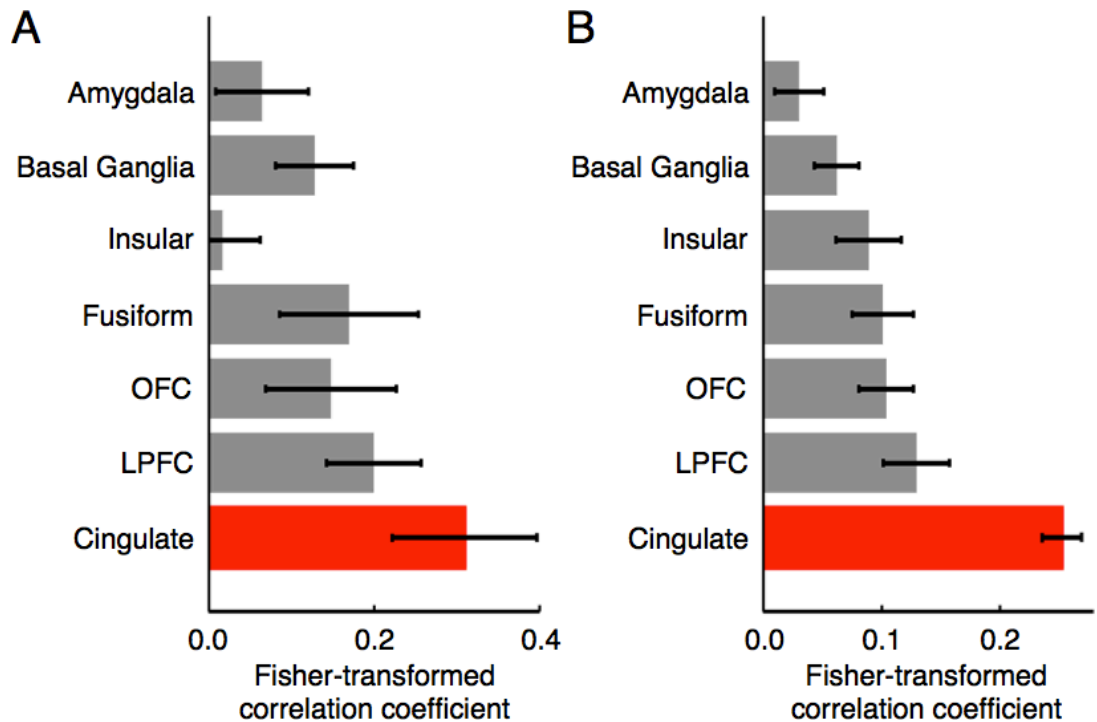

**S1 Fig. Comparison of decoding performance among seven regions in the fMRI decoder construction stage in the pilot and main experiments.** The pilot experiment (see *Pilot experiment* in Materials and Methods for details) was aimed to select a single region that would be used as a target for multi-voxel pattern induction in the main experiment. The selection was made based on comparison of performance of preference decoders among regions of interests (ROIs) that were implicated in facial preference [14-20]: the amygdala, basal ganglia, insular cortex, fusiform area, orbit frontal cortex (OFC), lateral prefrontal cortex (LPFC), and cingulate cortex (CC) (see *Pilot experiment* in Materials and Methods for the definition of the ROIs). Decoder's performance was defined as a correlation coefficient between subjects' behavioral preference ratings and the estimated rating by the decoder from activation patterns of a region (see *Pilot experiment* in Materials and Methods for details). (A) The mean (±

s.e.m) Fisher-transformed correlation coefficient in the pilot experiment ($N=3$). The CC showed the highest decoding performance. (B) The mean (± s.e.m) Fisher-transformed correlation coefficient in the main experiment ($N=24$). The decoder successfully predicted the subjects' behavioral preference ratings based on the fMRI datasets measured in the fMRI decoder construction stage in the basal ganglia (one-sample two-tailed t-test, $t_{23} = 2.39$, $P < 10^{-2}$; Bonferroni corrected), insular cortex ($t_{23} = 3.23$, $P < 10^{-2}$), fusiform area ($t_{23} = 3.89$, $P < 10^{-3}$, Bonferroni corrected), OFC ($t_{23} = 4.50$, $P < 10^{-3}$, Bonferroni corrected), LPFC ($t_{23} = 4.63$, $P < 10^{-3}$, Bonferroni corrected), CC ($t_{23} = 15.02$, $P < 10^{-4}$, Bonferroni corrected), but not in the amygdala ($t_{23} = 1.45$, $P = 0.16$). The mean performance in the CC was significantly higher than that in the other ROIs (paired two-tailed t-test, $t_{23} > 4.22$, $P < 10^{-3}$; Bonferroni corrected). The results of the main experiment (shown in B) showed the same tendency as in the pilot experiment (shown in A), indicating that activation patterns of the CC most accurately reflected subjects' behavioral preference ratings in the preference-rating task.

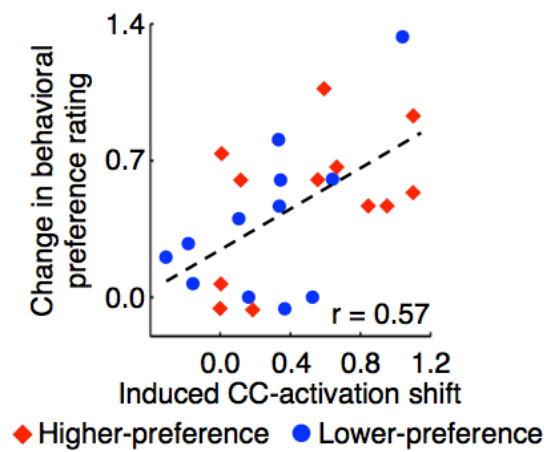

**S2 Fig. Significant correlation between the induced CC-activation shifts and the changes in the behavioral preference ratings.** The same scatter plot of the induced CC-activation shift during the 3-day induction stage vs. the change in subjects'

behavioral preference rating for the higher- (red diamonds; $N$=12) and lower- (blue circles; $N$=12) preference groups as in Fig 4, except that the sign of the data for the lower-preference group was reversed. The broken line indicates the ordinary least-square regression.

**S1 Data. Subjects' reports in the interview stage.** When subjects from the higher- and lower-preference groups were asked what they tried to do during the induction period of the induction stage to increase the size of the feedback disk, the most common reply was: "I tried various things since I had no idea of the correct way." Thus, we asked each subject to report the strategy that the subject thought was the most efficient to increase the size of the disk in detail. Details of the strategies that subjects reported were as follows. Note that these reports have been translated from Japanese into English:

Higher-preference group

Subject 1: "I tried to focus my attention on the fixation point at the center of the display."

Subject 2: "I tried to imagine that I am doing various techniques for gymnastics."

Subject 3: "I tried to remember contents of recent conversation with my friends in detail."

Subject 4: "I tried to imagine that I am singing a song."

Subject 5: "I tried to imagine that I am singing a song and dancing with a large number of people."

Subject 6: "I tried to focus my attention on the fixation point at the center of the display."

Subject 7: "I tried to imagine various colors."

Subject 8: "I tried to translate recent daily happenings into English."

Subject 9: "I tried to imagine that a face presented in the face period is moving."

Subject 10: "I tried to imagine the details of a building in my junior high school."

Subject 11: "I tried to remember items in a specific category (e.g., vegetable)."

Subject 12: "I tried to relax myself."

Lower-preference group

Subject 13: "I tried to relax myself."

Subject 14: "I tried to remember various characters in a famous video game."

Subject 15: "I tried to remember my friend's face which resembles a face presented in the face period."

Subject 16: "I tried to count numbers, remember various natural scenes, and relax and keep on switching these strategies once in a while."

Subject 17: "I tried to imagine that I am doing a back flip."

Subject 18: "I tried to perform difficult numerical calculations."

Subject 19: "I tried to relax myself."

Subject 20: "I tried to imagine various scenes in my daily life."

Subject 21: "I tried to imagine that I am listening to music."

Subject 22: "I tried to imagine that I am listening to music."

Subject 23: "I tried to remember detailed procedures in the experiment I conducted as an experimenter."

Subject 24: "I tried to remember contents of classes I have recently attended in my university."

No interview was conducted for the control group since they did not go through the induction stage with multi-voxel pattern induction.